\documentclass[12pt]{report}
\usepackage{graphicx}
\begin{document}
\voffset = -2.9cm
\hoffset = -1.3cm
\def\itm{\newline \makebox[8mm]{}}
\def\ls{\makebox[8mm]{}}
\def\fra#1#2{\frac{#1}{#2}}
\def\fr#1#2{#1/#2}
\def\frl#1#2{\mbox{\large $\frac{#1}{\rule[-0mm]{0mm}{3.15mm} #2}$}}
\def\frn#1#2{\mbox{\normalsize $\frac{#1}{\rule[-0mm]{0mm}{3.15mm} #2}$}}
\def\frm#1#2{\mbox{\normalsize $\frac{#1}{\rule[-0mm]{0mm}{2.85mm} #2}$}}
\def\frn#1#2{\mbox{\normalsize $\frac{#1}{\rule[-0mm]{0mm}{3.15mm} #2}$}}
\def\hs#1{\mbox{\hspace{#1}}}
\def\b{\begin{equation}}
\def\e{\end{equation}}
\def\arccot{\mbox{arccot}}
\vspace*{6mm}
\makebox[\textwidth][c]
{\large \bf{Closed timelike geodesics in a gas of cosmic
strings}}
\vspace{4mm} \newline
\makebox[\textwidth][c]
{\normalsize \O yvind Gr\o n$^{* \, \dag}$ and Steinar Johannesen$^*$}
\vspace{1mm} \newline
\makebox[\textwidth][c]
{\scriptsize $*$ Oslo University College, Department of Engineering,
P.O. box 4 St. Olavs plass, N-0130 Oslo, Norway}
\vspace{-5mm} \newline
\makebox[\textwidth][c]
{\scriptsize $\dag$ Institute of Physics, University of Oslo,
P.O. box 1048, N-0316 Oslo, Norway}
\vspace{-5mm} \newline
\makebox[\textwidth][c]
{\scriptsize E-mail: Oyvind.Gron@iu.hio.no and Steinar.Johannesen@iu.hio.no}
\vspace{3mm} \newline
{\bf \small Abstract}
{\small We find a class of solutions of Einstein's field equations
representing spacetime outside a spinning cosmic string surrounded by
vacuum energy and a rotating gas of non-spinning strings of limited radial
extension. Outside the region with vacuum energy and cosmic strings there
is empty space. We show that there exist closed timelike geodesics in this
spacetime.}
\vspace{7mm} \newline
{\bf 1. Introduction}
\begin{list}{}{\setlength{\leftmargin}{0mm}}
\item The question of the causal structure of spacetime is related
to the question of the existence of closed timelike curves (CTC).
One should therefore try to formulate as simple criteria as possible
for the existence of such curves. It would be particularly exciting if
a free particle could move so that it arrives back at the event it
started from. Such a particle must follow a closed timelike geodesic
(CTG).
\itm The existence of closed timelike curves has been investigated in
some cases. The first one to demonstrate the existence of a solution
of Einstein's field equations with closed timelike curves was
K.G\"{o}del [1]. He showed that such curves exist in a rotating
universe model with dust and negative vacuum energy. Already in 1937
W.J. van Stockum [2] found the solutions of Einstein's field equations
inside and outside a cylindrically symmetric rotating dust distribution.
F.J.Tipler [3] showed that the external solution contains CTC's.
J.R.Gott has shown that closed timelike curves exist in a spacetime with
a pair of moving cosmic strings [4]. An external time machine in
$(2 + 1)$-dimensional anti-de Sitter space has been analysed by
S.DeDeo and J.R.Gott [5].
Another example of a spacetime admitting closed timelike curves is
Bonnor's rotating dust metric which has recently been analysed by
P.Collas and D.Klein [6].
\itm Only a few known solutions contain closed timelike geodesics.
The first one was presented by E.D.Soares [7] in 1979.
He found a class of cosmological solutions of the Einstein-Maxwell
equations with rotating dust and electromagnetic fields. For a subclass
of these models, the topology of the spacetime manifold is
$\mbox{\bf $S$}^3 \times \mbox{\bf $R$}$, and the timelike geodesic
curves followed by the dust particles are closed. In this case
the existence of the closed timelike geodesics depends upon the
nontrivial topology of the considered spacetime.
\itm The first one to have demonstrated the existence of closed timelike
geodesics in $\mbox{\bf $R$}^4 \,$ is B.R.Steadman [8]. He has shown that
such curves exist in the exterior van Stockum spacetime. It has also been
shown by W.B.Bonnor and B.R.Steadman [9] that a spacetime with two
spinning particles, each one with a magnetic moment and a mass equal to
their charge, permit special cases in which there exist closed timelike
geodesics. V.M.Rosa and P.L.Letelier [10] have shown that there exist
regions where these curves are stable.
\itm Recently V.M.Rosa and P.L.Letelier have constructed a new solution
of Einstein's field equations with a spacetime containing a
black hole pierced by a spinning cosmic string [11].
\itm In the present article we find a class of new solutions of
Einstein's field equations describing a spacetime with closed
timelike geodesics. This spacetime is time independent and cylindrically
symmetric. It is filled with a gas of limited radial extension
which consists of non-spinning strings parallel to the symmetry axis
and vacuum energy, and there is a spinning cosmic string along the axis.
Spacetime in the gas is matched continuously to the external metric
in empty space outside a spinning cosmic string [12].
\end{list}
\vspace{6mm}
{\bf 2. Physical properties of the reference frame}
\vspace{-5mm} \newline
\begin{list}{}{\setlength{\leftmargin}{0mm}}
\item We shall consider curved spacetime with axial symmetry.
The metric is assumed to be stationary. The line element may then be
written [13]
\b
ds^2 = -f dt^2 + 2k \hs{0.3mm} dt \, d\phi + l \hs{0.5mm} d \phi^2
+ e^{\mu} (dr^2 + dz^2) \mbox{ ,}
\e
where $f$, $k$, $l$ and $\mu$ are functions of $r$ and $z$, and we have
used units so that the velocity of light $c = 1$. It may be noted
that Lorentz invariance in the $z$-direction implies that $f = e^{\mu}$.
The coordinates $r,\phi,z$ are comoving in a reference frame $R$.
\itm Rindler and Perlick [14] have deduced a formula for the
four-acceleration $A_{\mu}$ and the vorticity $\Omega$ of the
four-velocity field of the reference particles in $R$ relative to
the local compass of inertia. Applied to the line element (1) one
obtains [15]
\b
A_{\mu} = (0,\frn{f_{,r}}{2f},0,0)
\e
and
\b
\Omega = \frn{e^{- \mu / 2} \hs{0.5mm} f}{2 \hs{0.5mm} D}
\left( \frn{k}{f} \right)_{\hs{-0.5mm} ,r}
\e
where
\b
D^2 = fl + k^2 > 0 \mbox{ .}
\e
The condition $fl + k^2 > 0$ comes from the signature of the spacetime
metric. Here $\Omega$ represents the local rate of rotation of the
reference frame $R \,$, i.e. the rotation with respect to a "compass of
inertia".
A more general formula, valid also for time dependent metrics has been
deduced by Weissenhoff [16]. When $k/f$ is constant, the vorticity is
vanishing. Since the metric is time independent, the distances between
fixed reference points are also independent of time. In this sense the
reference system is rigid. Thus, when $k/f$ is constant, the swinging
plane of the Foucault pendulum, representing a local compass of inertia,
will have a fixed direction in $R$.
\itm The coordinate clocks showing the time $t$ are at rest in
$R$, but the existence of the product term $dt \, d\phi$ means that
these clocks are not Einstein synchronized around a closed curve when
$k \ne 0 \,$ [17].
\itm In order to investigate the kinematics of the frame $R$, we consider
a particle moving in the a plane $z = \mbox{constant} \,$. The Lagrange
function of the particle is
\b
L = - \frn{1}{2} f \dot{t}^2 + k \dot{t} \dot{\phi} +
\frn{1}{2} l \dot{\phi}^2 + \frn{1}{2} e^{\mu} \dot{r}^2
\e
The conserved angular momentum of the particle is
\b
p_{\phi} = \frl{\partial L}{\partial \dot{\phi}} =
l \dot{\phi} + k \dot{t}
\e
where the dot denotes differentiation with respect to the proper time
of the particle.
\itm In the case $l > 0$, the coordinate basis vector $\vec{e}_{\phi}$
is spacelike, and we can define an angular velocity
\b
\omega = \frl{d \phi}{dt} = \frl{\dot{\phi}}{\dot{t}}
\e
giving
\b
p_{\phi} = (l \omega + k) \dot{t}
\e
A local inertial reference frame is a freely falling, non-rotating frame
having $p_{\phi} = 0$. Hence the inertial frames have an angular velocity
\b
\omega = - \frl{k}{l}
\e
in the reference frame $R$. This is an expression of the inertial
dragging effect. Hence, in the case $l > 0$, the presence of the
$d \phi dt$-term in the line element (1) is due to the rotation of the
reference frame $R$ relative to the inertial frames. We will now argue
that this interpretation is not valid in the case $l < 0$.
\itm The line element (1) may be written
\b
ds^2 = - \frm{D^2}{l} \hs{0.8mm} d t^2
+ \hs{0.0mm} l \left( d\phi + \frm{k}{l} \hs{0.5mm} dt \right)^2
+ e^{\mu} (dr^2 + dz^2)
\mbox{ .}
\e
This form of the line element is obtained by replacing the coordinate
basis vector field $(\vec{e_t},\vec{e}_{\phi},\vec{e}_r,\vec{e}_z)$
by the orthogonal basis vector field
$(\vec{e}_t - (k / l) \vec{e}_{\phi}, \vec{e}_{\phi},\vec{e}_r,\vec{e}_z)$.
The vector $\vec{e}_t - (k / l) \vec{e}_{\phi}$ is timelike and
$\vec{e}_{\phi}$, $\vec{e}_r$ and $\vec{e}_z$ are spacelike when $l > 0$,
which leads to the usual intepretation  of $\omega$ as an angular velocity
as given in eq.(9). However, $\vec{e}_{\phi}$ is timelike and
$\vec{e}_t - (k / l) \vec{e}_{\phi}$, $\vec{e}_r$ and $\vec{e}_z$ are
spacelike when $l < 0$, making this interpretation impossible.
\itm The line element can also be written in the form
\b
ds^2 = - f \hs{0.0mm} \left( dt - \frm{k}{f} \hs{0.5mm} d \phi
\right)^2 + \frm{D^2}{f} \hs{0.8mm} d \phi^2 +
e^{\mu} (dr^2 + dz^2)
\e
when $l > -k^2$, permitting both positive and negative values of $l$.
This is obtained by replacing the coordinate basis vector field
$(\vec{e_t},\vec{e}_{\phi},\vec{e}_r,\vec{e}_z)$
with the orthogonal basis vector field
$(\vec{e}_t, \vec{e}_{\phi} + (k / f) \vec{e}_t, \vec{e}_r,\vec{e}_z)$.
The $3$-space spanned by the spacelike basis vectors
$(\vec{e}_{\phi} + (k / f) \vec{e}_t,\vec{e}_r,\vec{e}_z)$ represents
the simultaneity space orthogonal to the timelike basis vector
$\vec{e}_t$. Hence the form (11) of the line element separates out
the $3$-space orthogonal to the four-velocity of the reference particles.
\itm The expression $dt - (k / f) d \phi$ is an exact differential when
$k / f$ is constant. In this case one can introduce a new time coordinate
\b
\hat{t} = t - \frn{k}{f} \hs{0.5mm} \phi
\mbox{ ,}
\e
and the line element takes the form
\b
ds^2 = - f d\hat{t}^2 + \frn{D^2}{f} \hs{0.8mm} d \phi^2 +
e^{\mu} (dr^2 + dz^2) \mbox{ .}
\e
The clocks showing $\hat{t}$ are locally Einstein synchronized.
Hence the coordinate clocks showing $t$ are not Einstein synchronized.
\itm In the case $l > 0$, the reference frame is rotating. Then the
Einstein synchronised clocks showing $\hat{t}$ have a gap when one goes
around a circle about the axis. In this case the time $\hat{t}$ is not
well globally defined due to the mentioned gap.
\itm As shown above, when $l < 0$ the reference system is not rotating.
In this case one may Einstein synchronize clocks around a circle. Then
the time $\hat{t}$ given by eq.(12) is well defined globally. Hence,
although the form (11) of the line element is valid both for $l > 0$
and $l < 0$, it introduces this well defined coordinate $\hat{t}$ only
when $l < 0$.
\itm Locally the situation is then similar to that of a person travelling
westwards around the equator in an airplane so that the local time is
constant on the Earth at the position of the airplane. This corresponds
to keeping the time $t$ constant. However, the proper time of the passenger
increases, corresponding to an increase of the time $\hat{t}$ as given
in eq.(12), when $\phi$ increases. Going around a circle, there is a gap
of the time $t$ on the Earth at the date line.
\itm However, spacetime may be different. In the present article we
shall consider spacetime outside a spinning cosmic string. Then there
is no time gap in $t$ when one goes around the circle.
%
\end{list}
\vspace{6mm}
{\bf 3. General criteria for closed timelike geodesics}
\vspace{-5mm} \newline
\begin{list}{}{\setlength{\leftmargin}{0mm}}
\item Let us consider a circular timelike curve in the plane $z = constant$.
The $4$-velocity identity applied to the tangent vector field of the
curve takes the form
\b
l \dot{\phi}^2 + 2 k \hs{0.5mm} \dot{t} \dot{\phi} - f \dot{t}^2 = -1
\mbox{ ,}
\e
where we have used units so that the velocity of light $c = 1 \,$,
and the dot means the derivative with respect to proper time $\tau$.
If $\dot{t} = 0$, the curve is closed in spacetime. This condition
seems to mean that a particle moving along such a curve is everywhere
on the curve at a fixed point of time, i.e. that it has an infinitely
great velocity. One may wonder if this is in conflict with the special
theory of relativity. However, this infinitely great velocity is measured
with clocks that are not Einstein synchronized. As measured by
Einstein synchronized clocks, the velocity of the particle is less than
the velocity of light. Hence its world line is inside the light cone,
which it what it means to say that it is timelike.
For such a curve the $4$-velocity identity reduces to
\b
l \dot{\phi}^2 = -1 \mbox{ .}
\e
Hence the metric component $l < 0$ on the curve.
\itm We shall now consider closed {\it geodesic} curves in a plane
$z = \mbox{constant}$. The radial Lagrange equation is
\b
\frac{d}{d \tau} \left( \frac{\partial L}{\partial \dot{r}} \right) =
\frac{\partial L}{\partial r}
\e
In the case of a circular geodesic in the $z = constant$ plane,
the Lagrangian of a free particle is
\b
L = - \frn{1}{2} f \dot{t}^2 + k \dot{t} \dot{\phi} +
\frn{1}{2} l \dot{\phi}^2
\mbox{ ,}
\e
and the equation takes the form
\b
l_{,r} \dot{\phi}^2 + 2k_{,r} \dot{t} \dot{\phi}
- f_{,r} \dot{t}^2 = 0 \mbox{ .}
\e
The condition that the curve shall be closed is $\dot{t} = 0$.
From eqs.(15) and (18) we then have the following two conditions
which must be fullfilled on a closed circular geodesic curve with
radius $r = r_2$ :
\b
l(r_2) < 0
\mbox{\hspace{3mm} , \hspace{3mm}}
l_{,r}(r_2) = 0 \mbox{ .}
\e
Similarly, from the $z$-component of the geodesic equation, we
obtain the condition
\b
l_{,z}(r_2) = 0 \mbox{ .}
\e
The first of the conditions (19) comes from the requirement that the
curve shall be closed in spacetime, the second as well as (20), that
it is also geodesic.
\end{list}
\vspace{6mm}
{\bf 4. Applications to some known metrics}
\vspace{-5mm} \newline
\begin{list}{}{\setlength{\leftmargin}{0mm}}
\item
{\it 4a. The Bonnor metric}
\vspace{1.5mm} \newline
This metric represents spacetime in a rotating cloud of dust.
It has earlier been investigated by Collas and Klein [4]. Let us
apply the conditions (19) and (20) to Bonnor's metric, having
\b
l = r^2 - 4 h^2 r^4 / (r^2 + z^2)^3 \mbox{ .}
\e
In this case the condition (20) takes the form
\b
l_{,z}(r_2) = 24 h^2 r_2^4 z / (r_2^2 + z^2)^4 = 0
\e
which requires that $z = 0$. Hence timelike circular closed
geodesics in the hyperplane of $z = constant$, with center on the
$z$-axis, can only exist in the plane $z = 0$. The condition
$\, l_{,r}(r_2) = 0 \,$ takes the form $r_2^4 = - 4 h^2$,
showing that there are no closed, circular geodesics in the plane $z = 0$.
This shows that the Bonnor metric does not permit the existence of
such closed, timelike geodesics. This provides a simple proof of a
result that was originally found by Collas and Klein [4].
\vspace{5mm} \newline
{\it 4b. The Van Stockum spacetime}
\vspace{1.5mm} \newline
We shall apply our criteria (19) to a van Stockum metric [2] which
has earlier been analysed by B.R.Steadman [6], describing an empty
space outside a rotating dust cylinder. In this case
the relevant metric function (in our notation) is
\b
l = \frac{R r \hs{0.3mm} \sin (3 \epsilon + \log(r / R) \tan \epsilon)}
{\sin \epsilon + \sin (3 \epsilon)}
\e
where
\b
\tan \epsilon = (4 a^2 R^2 - 1)^{1/2}
\mbox{\hspace{3mm} , \hspace{3mm}}
\frl{1}{2} < aR < 1 \mbox{ .}
\e
The constant $a$ is the spin per mass of the rotating dust.
\itm The condition $l_{,r}(r_2) = 0$ for a CTG now takes the form
\b
r_2 = R \hs{0.9mm} \exp \hs{0.2mm} ((k \pi - 4 \epsilon) \cot \epsilon)
\e
where $k$ is an integer. Inserting this into the expression for
$l$ gives
\b
l(r_2) = (-1)^{k + 1} \hs{0.5mm} \frac{\sin \epsilon}
{\sin \epsilon + \sin (3 \epsilon)} \hs{1.0mm} r_2 R \mbox{ .}
\e
Using equation (24) this can be written
\b
l(r_2) = (-1)^{k + 1} a^2 R^3 r_2 \mbox{ .}
\e
In addition, the condition $l(r_2) < 0$ must be fullfilled. This requires
that $k = 2n$ where $n$ is an integer. Hence there exists closed timelike
circular geodesics with radii
\b
r_2 = R \hs{0.9mm} \exp \hs{0.2mm} (2(n \pi - 2 \epsilon) \cot \epsilon)
\e
where $n$ is an integer. Since the present metric is a solution
of the field equations in empty space outside the dust distribution,
the radii must satisfy $r_2 > R$. Hence $n \pi > 2 \epsilon$ which
demands that $n$ is a positive integer. We have thereby given a simple
demonstration of a result which was first found by Steadman [6].
\vspace{5mm} \newline
{\it 4c. The G\"{o}del universe}
\vspace{1.5mm} \newline
We shall here apply our criteria to the G\"{o}del universe model [1].
This is a rotating universe filled with dust and vacuum energy. The
density of the vacuum energy is negative and has half the value of
the density of the dust.
Barrow and Tsagas [18] have recently analysed the stability and dynamics
of the G\"{o}del universe and also discussed the existence of closed
timelike curves in this universe model. The metric function
$g_{\phi \phi}$ is
\b
l = 4 a^2 \left[ \sinh^2 \left(\frl{r}{2a} \right) -
\sinh^4 \left(\frl{r}{2a} \right) \right] \mbox{ ,}
\e
where $a^2 = 1/\kappa \mu$ and $\mu$ is the density of the dust.
Then $l_{, r} = 0$ leads to
\b
\sinh \left( \frl{r_2}{2a} \right) = \frl{1}{\sqrt{2}} \mbox{ .}
\e
Inserting this into eq.(29) we get $\, l = l_0 = a^2 > 0 \,$. Hence
closed timelike circular geodesics do not exist in this universe model.
However, there is a region with closed timelike non-geodesic curves.
In this region $l < 0 \,$, which corresponds to
$\sinh (r / 2a) > 1$, i.e.
$r > 2a \log (1 + \sqrt{2}) \,$.
\end{list}
\newpage
{\bf 5. New solutions with closed timelike geodesics}
\vspace{-5mm} \newline
\begin{list}{}{\setlength{\leftmargin}{0mm}}
\item We shall consider a cylindrically symmetric space outside a
spinning cosmic string filled with vacuum energy and a non-rotating gas
of non-spinning strings oriented parallel to the axis. The vacuum energy
and the gas of strings have limited extension in the radial direction.
We shall find the metric inside the gas by solving Einstein's field
equations, demanding that it matches continuously at the outer boundary
of the mass distribution to the metric [12] in empty space outside a
spinning cosmic string.
\itm We now assume that $f = e^{\mu} = 1$ and that $k$ is constant.
According to eq.(3) this means that the vorticity of the four-velocity
field of the reference particles vanishes. From eq.(2) it also follows
that the acceleration of gravity vanishes, so that the gas of strings
is in static equilibrium. We shall construct solutions with $l < 0$ in
the region with cosmic strings. Hence the reference frame is non-rotating
in this region. The line element then takes the form
\b
ds^2 = - \hs{0.0mm} \left( dt - k \hs{0.5mm} d \phi
\right)^2 + D^2 \hs{0.8mm} d \phi^2 + dr^2 + dz^2 \mbox{ .}
\e
This form of the metric with $k = \mbox{constant}$ has recently been
discussed by Cooperstock \& Tieu [19] and Culetu [20]. In this case it
is possible to introduce a new time coordinate $\hat{t} = t - k \phi$
so that the line element takes the form
\b
ds^2 = - d\hat{t}^2 + D^2 \hs{0.8mm} d \phi^2 + dr^2 + dz^2 \mbox{ .}
\e
The 3-space of the gas is given by the last three terms in eq.(32).
\itm Going around a curve from $\phi = 0$ to $\phi = 2 \pi$, either $t$
or $\hat{t}$ (or both) have to make a gap, and in general there seems
to be lacking a criterion to decide which one. It may be a question
of defining the topology of the spacetime which is considered.
\itm In the present work we assume that the spacetime outside the gas
which surrounds the string is that of empty space outside a spinning
cosmic string [12]. Then the original time coordinate $t$ does not have
a gap, while $\hat{t}$ has a gap equal to $2 \pi k$.
\itm The nonzero contravariant components of the energy-momentum tensor
in the gas of strings are
\b
T^{tt}_g = \rho_g
\mbox{\hspace{3mm} , \hspace{3mm}}
T^{zz}_g = p_g
\e
We assume that the vacuum energy has a density $\rho_{\Lambda} > 0$. It is
also assumed to have a pressure $p_{\Lambda}$ only in the axial direction.
The nonzero contravariant components of the energy-momentum tensor for the
vacuum are
\b
T^{tt}_{\Lambda} = \rho_{\Lambda}
\mbox{\hspace{3mm} , \hspace{3mm}}
T^{zz}_{\Lambda} = p_{\Lambda}
\e
where $p_{\Lambda} = -{\rho}_{\Lambda}$ is constant.
There is no energy current since the gas is at rest in the
reference frame.
\itm Calculating the Einstein tensor for the line element (31),
we find that the $tt-$ and $zz$-components of Einstein's field
equations take the form
\b
-\frl{D''}{D} = \kappa (\rho_g - \rho_{\Lambda} g^{tt})
= \kappa (\rho_g + \rho_{\Lambda} l / D^2)
\e
and
\b
\frl{D''}{D} = \kappa (p_g + p_{\Lambda} g^{zz})
= \kappa (p_g + p_{\Lambda})
\e
where $\kappa$ is Einstein's gravitational constant. From these
equations we obtain
\b
{\rho}_g = - (p_g + \frn{k^2}{D^2} p_{\Lambda})
= \frn{k^2}{D^2} {\rho}_{\Lambda} - p_g
\e
Note that in the present case the usual coordinate condition
$D(r) = r$ does not permit a non vanishing density, and can therefore
not be applied in the space inside the gas.
\itm The metric function $l$ depends upon the density as a
function of $r$ which may be chosen freely. Furthermore
\b
l(r) = D(r)^2 - k^2 \mbox{ .}
\e
For a bounded function $D(r)$ the condition $l(r) < 0$ may be
fullfilled by choosing $k$ sufficiently large. Hence there are closed
timelike curves in this spacetime. However, these curves will in general
not be geodesics. If they shall be geodesics, the additional requirement,
$l'(r) = 2 D(r) D'(r) = 0$ must be met. In order not to have a degenerate
metric, we must assume that $D(r) \ne 0$. Then the condition $l'(r) = 0$
is equivalent to the condition $D'(r) = 0$, showing that the two
conditions (19) for a closed circular geodesic curve may be fullfilled.
\itm In the situation we consider, there is a spinning cosmic string
along the $z$-axis. The non vanishing energy-momentum tensor components
of the string are given by
\b
\sqrt{g} \hs{1.9mm} T^{tt} = - \sqrt{g} \hs{1.9mm} T^{zz}
= {\lambda}_{\hs{0.5mm} cs} \hs{0.5mm} \delta^{(2)}(\mbox{\bf $r$})
\mbox{\hspace{3mm} , \hspace{3mm}}
\sqrt{g} \hs{1.9mm} T^{t \phi} =
\frn{1}{2} \hs{0.9mm} J \hs{0.9mm} \frl{\partial}{\partial r} \hs{0.5mm}
\delta^{(2)}(\mbox{\bf $r$})
\e
where ${\lambda}_{\hs{0.5mm} cs}$ and $J$ are the mass and the angular
momentum per unit length of the string [21].
The line element in empty space outside this string is [12]
\b
ds^2 = - \hs{0.0mm} \left( dt - J / {\lambda}_{\hs{0.5mm} 0} \hs{0.5mm}
d \phi \right)^2
+ (1 - \lambda / {\lambda}_{\hs{0.5mm} 0})^2 \hs{0.8mm} r^2 d \phi^2
+ dr^2 + dz^2
\e
where $\lambda$ is the total mass per unit length of the spinning string
on the z-axis and the gas and vacuum outside it. Furthermore
${\lambda}_{\hs{0.5mm} 0}
= \hat{\lambda} / 4$ where $\hat{\lambda} =
c^2 / \hs{0.3mm} G = 1.3 \cdot 10^{27} \hs{0.5mm} \mbox{kg} / \mbox{m}$
is the gravitational line density. This solution has
\b
l(r) = (1 - \lambda / {\lambda}_{\hs{0.5mm} 0})^2 \hs{0.5mm} r^2
- (J / {\lambda}_{\hs{0.5mm} 0})^2
\e
\vspace{-5mm}
\itm Assume that the exterior boundary of the gas is at $r = r_1$. Then
the components of the metric tensor and their derivatives must be
continuous at $r = r_1$. Hence
\b
k = J / {\lambda}_{\hs{0.5mm} 0}
\e
\b
D(r_1) = (1 - \lambda / {\lambda}_{\hs{0.5mm} 0}) \hs{0.5mm} r_1
\e
\b
D'(r_1) = 1 - \lambda / {\lambda}_{\hs{0.5mm} 0}
\e
\vspace{-3mm}
\itm We see that $D''$ and $D$ must have equal signs. Otherwise the graph
of $D(r)$ curves towards the $r$-axis. In this case $D(r) = 0$ for a value
of $r$ between each timelike closed geodesic where $D'(r) = 0$, and
between $r = r_1$ and the geodesic of maximum $r$.
These values of $r$ correspond to cylindrical surfaces in spacetime where
the metric is degenerate. In order to obtain a connected spacetime, we
must therefore assume that $D'' / D > 0$. According to eqs.(35) and (36)
this implies that
\b
{\rho}_g < - \frn{l}{D^2} {\rho}_{\Lambda}
\e
and
\b
p_g > - p_{\Lambda}
\e
Note that when $l < 0$, eq.(45) permits a positive mass density of the
gas of strings. Also, since $p_{\Lambda} = - {\rho}_{\Lambda}$, it follows
from eq.(46) that $p_g > 0$. Hence the pressure and the density of the
gas of cosmic strings have the same sign. This is unusual. In the absence
of the vacuum energy the cosmic gas would, according to eq.(37), obey the
equation of state $p_g = -{\rho}_g$. In the present model, however,
equation (37) tells that the equation of state of the gas of cosmic
strings is modified by the vacuum energy. Also, there are no cosmic
strings in the external region where there is no vacuum energy.
\itm In the present case the sign of $D(r)$ cannot change, and we can
therefore assume that $D(r) > 0$ for every value of $r > 0 \,$.
This means that the line density of the string must fullfill
$\lambda < {\lambda}_0 \,$ which secures the possibility of a weak
field approximation.
\itm By choosing
\b
J = ({\lambda}_0 - \lambda) r_1
\mbox{ ,}
\e
we obtain $l(r_1) = 0$ from eq.(41). With this choice of the angular
momentum per unit length of the string, we have that $l(r) > 0$ for
$r > r_1$. According to our analysis in section 2 this means that the
reference frame $R$ is rotating in this region. From eqs.(9) and (3)
it follows that the angular velocity of the reference frame decreases
for increasing $r$ in such a way that the vorticity of the four-velocity
field of the reference particles vanishes also in this region.
We consider solutions with the property that in the region
$0 < r \le r_1$, the function $D(r)$ has a maximum at $r = r_1$,
ensuring that $l < 0$ in the gas of strings. This means that the
reference frame is non-rotating in the entire region with vacuum
energy and strings.
\itm We shall now discuss two different solutions with these properties.
\vspace{2mm} \newline
(i) The pressure $p_g$ of the gas is constant inside a cylindrical surface
with radius $r_1$. The total pressure in the combined system of strings
and vacuum energy is $p_0 = p_g + p_{\Lambda}$. Solving the field
equations then gives
\b
D(r) = \beta \cosh \left( \frl{r - r_1}{r_0} + \alpha \right)
\e
for $r \le r_1$, where $\beta > 0$ and $\alpha$ are integration constants.
In the region $r \le r_1$, the function $D(r)$ has a maximum for $r = r_1$.
Thus $l(r) < 0$ for $r < r_1$.
Furthermore $r_0 = 1 / \sqrt{\kappa p_0} \,$, which may be written
$r_0 = 0.75 (p_1 / p_0)^{1/2} 10^{13} m
= 7.9 (p_1 / p_0)^{1/2} 10^{-4} l.y.$, where
$p_1 = - 1 N / m^3$.
Eq.(37) shows that the density of the gas is positive when
\b
p_g < \frn{k^2}{D^2} {\rho}_{\Lambda}
\e
Applying the conditions (43) and (44) to the solution (48) leads to
\b
\beta \cosh (\alpha) = (1 - \lambda / {\lambda}_{\hs{0.5mm} 0})
\hs{0.5mm} r_1
\mbox{\hspace{3mm} , \hspace{3mm}}
\beta \sinh (\alpha) = (1 - \lambda / {\lambda}_{\hs{0.5mm} 0})
\hs{0.5mm} r_0
\e
showing that $r_1 > r_0$. Hence
\b
\beta = (1 - \lambda / {\lambda}_{\hs{0.5mm} 0}) \hs{0.5mm}
\sqrt{r_1^2 - r_0^2}
\mbox{\hspace{3mm} , \hspace{3mm}}
\alpha = \mbox{artanh} \left( \frl{r_0}{r_1} \right)
\e
It may be noted that if the gas is removed, corresponding to
$r_0 \rightarrow \infty$, we get $\lambda \rightarrow
{\lambda}_{\hs{0.5mm} cs}$ giving the spacetime outside the string [12].
\itm The condition $D'(r_2) = 0$ for a closed circular timelike
geodesic with radius $r_2$ leads to
\b
r_2 = r_1 - r_0 \hs{0.5mm} \alpha
\e
In order that $r_2 > 0 \,$, the radius $r_1$ of the gas must fullfill
\b
r_1 > r_0 \hs{0.5mm} \alpha
\e
which may be written
\b
\frl{r_1}{r_0} \hs{0.5mm} \tanh \left( \frl{r_1}{r_0} \right) > 1
\mbox{ .}
\e
This means approximately that $r_1 > 1.2 \hs{0.9mm} r_0 \,$, which
implies that $p_0 < - 1.4 / {\kappa r_1^2}$. This shows that
there is no closed timelike geodesic without the gas surrounding
the spinning string. In this model there is a tension in the
$z$-direction, not a pressure, since $p_0 < 0$.
\vspace{-2mm} \newline
\itm (ii) The total pressure $p = p_g + p_{\Lambda}$ of the gas and
vacuum energy is assumed to be
\b
p(r) = \frl{2}{\kappa \hs{0.5mm} [(r - a)^2 + b^2]}
\e
where the constant $b$ represents a length corresponding to the total
pressure of the gas and vacuum at the radius $r = a$,
\b
b^2 = \frl{2}{\kappa p(a)}
\mbox{ .}
\e
Note that $p(r) > 0$ in accordance with eq.(46). We assume that the
vacuum energy and the gas extends to a radius
\b
r_1 = \sqrt{a^2 + b^2}
\mbox{ .}
\e
Outside this radius the spacetime is described by the metric (40).
\itm The field equation (36) now takes the form
\b
D'' - \frl{2}{(r - a)^2 + b^2} \hs{1.2mm} D = 0
\e
Introducing
\b
u = \frl{D}{(r - a)^2 + b^2}
\e
this equation can be written
\b
\{ [(r - a)^2 + b^2]^2 u' \}' = 0
\e
which implies that
\b
[(r - a)^2 + b^2]^2 u' = c_1
\e
The general solution of this equation is
\b
u = \frl{c_1}{2 b^3} \left[ \frl{b(r - a)}{(r - a)^2 + b^2} +
\arctan \frl{r - a}{b} \right] + c_2
\e
where $c_1$ and $c_2$ are integration constants.
Using eqs.(59) and (57), the condition (43) leads to
\b
2 (r_1 - a) u(r_1) = 1 - \lambda / {\lambda}_{\hs{0.5mm} 0}
\e
Using eqs.(59) and (61), the condition (44) gives
\b
2 (r_1 - a) u(r_1) + \frl{c_1}{(r_1 - a)^2 + b^2} =
1 - \lambda / {\lambda}_{\hs{0.5mm} 0}
\e
Eqs.(63) and (64) imply that
\b
c_1 = 0
\mbox{\hspace{3mm} , \hspace{3mm}}
c_2 = \frl{1 - \lambda / {\lambda}_{\hs{0.5mm} 0}}{2 (r_1 - a)}
\e
so that
\b
D(r) = c_2 [(r - a)^2 + b^2]
\e
In order that $D(r) > 0$, the condition $\lambda < {\lambda}_0$ must be
fullfilled since $r_1 > a$.
\itm In order for the reference frame to be non-rotating in the entire
region with vacuum energy and strings, the condition $l(r) < 0$ must
be satisfied for $0 < r \le r_1$. Hence, in this region the function $D(r)$
describing a parabola must have a maximum for $r = r_1$, which demands
that $D(0) \le D(r_1)$ giving $r_1 \ge 2a$. Inserting the expression
(57) for $r_1$ this condition may also be expressed as
$b \ge \sqrt{3} \hs{0.5mm} a$.
\itm For the present solution the conditions for the existence of closed
timelike geodesics give
\b
r_2 = a
\mbox{\hspace{3mm} , \hspace{3mm}}
J > D(a) {\lambda}_0
\mbox{ .}
\e
Thus, the constant $a$ represents the radius of the closed timelike
geodesics. With $J$ as given in eq.(47), and using eq.(42) and (38),
we have $J = D(r_1) {\lambda}_0$. Hence, the inequality (67) may be
written $D(r_1) > D(a)$, which is clearly satisfied for the parabola
described by eq.(66).
\end{list}
\vspace{6mm}
{\bf 6. Isometric embedding}
\vspace{-5mm} \newline
\begin{list}{}{\setlength{\leftmargin}{0mm}}
\item In this section we shall find the embedding surface for the
3-space $\hat{t} = constant$ in case (ii). Writing the line element
of the spacetime in the gas of cosmic strings in the form (11), we obtain
\b
ds^2 = - d\hat{t}^2 + c_2^2 [(r-a)^2 + b^2]^2 d \phi^2 + dr^2 + dz^2
\e
where $\hat{t} = t - (J / {\lambda}_0) \phi$ and $c_2$ is given in
eq.(65). It may be noted that this metric has a strange property.
Consider a circle in the plane $z = const$ having its center on the axis.
The quotient between the length of this circle and the diameter tends to
infinity in the limit $r \rightarrow 0 \,$.
However, $r = 0$ is not part of the spacetime described
by the line element (68), since there is a cosmic string along the
$z$-axis. The spatial part of this line element represents the space $S$
defined by simultaneity on Einstein synchronized clocks at rest in the
reference frame $R$.
\itm We shall now show that the $r \phi$ - surface with $z = constant$
can be isometrically embedded in Euclidean $3$-space. The line element
of this $3$-space in cylinder coordinates $(\hat{r}, \phi, \hat{z})$ is
\b
dl^2 = d \hat{r}^2 + \hat{r}^2 d \phi^2 + d \hat{z}^2
\e
The line element of the $r \phi$ - surface is
\b
dl_{r \phi}^2 = dr^2 + c_2^2 [(r-a)^2 + b^2]^2 d \phi^2
= dr^2 + \hat{r}(r)^2 d \phi^2
\e
where we have introduced the function
\b
\hat{r}(r) = c_2 [(r-a)^2 + b^2]
\e
The line element (70) can be reorganized as follows
\b
dl_{r \phi}^2 = d \hat{r}^2 + \hat{r}(r)^2 d \phi^2 +
\left[ 1 - \left( \frn{d \hat{r}}{dr} \right)^2 \right] dr^2
\e
This will coincide with the line element in the $3$-space given by eq.(66)
if we choose
\b
d \hat{z} = \sqrt{1 - \left( \frn{d \hat{r}}{dr} \right)^2}  dr
\e
Hence
\b
\hat{z} = \int_a^r \hs{-0.5mm}
\sqrt{1 - 4 c_2^2 (\overline{r}-a)^2} d\overline{r}
\e
which leads to
\b
\hat{z} = \frn{1}{2} \hs{0.9mm} (r - a) \hs{0.5mm}
\sqrt{1 - 4 c_2^2 (r-a)^2} + \frn{1}{4 c_2} \hs{0.5mm}
\arcsin \left\{ 2 c_2 (r-a) \right\}
\e
where $0 < r \le r_1$. The embedding is thus given by (75) together with
\b
x = c_2 [(r-a)^2 + b^2] \hs{0.5mm} \cos \phi
\mbox{\hspace{0.5mm} , \hspace{1mm}}
y = c_2 [(r-a)^2 + b^2] \hs{0.5mm} \sin \phi
\e
The embedding of the external space, given by the spatial part of the
line element (40), has
\b
\hat{z} = \hat{z} (r_1) + \sqrt{\frl{\lambda}{{\lambda}_{\hs{0.5mm} 0}}
\left( 2 - \frl{\lambda}{{\lambda}_{\hs{0.5mm} 0}} \right)}
\hs{1.5mm} (r - r_1)
\e
for $r \ge r_1$. Hence, it is described by a part of a cone in the
embedding diagram. The embedded surface is shown in Fig.1.
\vspace{55mm} \newline
\begin{picture}(50,0)(-96,-167)
\qbezier(168.1132, -138.0271)(166.0892, -136.4498)(164.2748, -134.1226)
\qbezier( 86.6038, -138.0271)( 88.6278, -136.4498)( 90.4422, -134.1226)
\qbezier(164.2748, -134.1226)(162.6282, -132.0106)(161.3396, -129.4309)
\qbezier( 90.4422, -134.1226)( 92.0888, -132.0106)( 93.3774, -129.4309)
\qbezier(161.3396, -129.4309)(160.1154, -126.9803)(159.3075, -124.2388)
\qbezier( 93.3774, -129.4309)( 94.6016, -126.9803)( 95.4095, -124.2388)
\qbezier(159.3075, -124.2388)(158.5215, -121.5718)(158.1785, -118.7522)
\qbezier( 95.4095, -124.2388)( 96.1955, -121.5718)( 96.5384, -118.7522)
\qbezier(158.1785, -118.7522)(157.8393, -115.9631)(157.9528, -113.1441)
\qbezier( 96.5384, -118.7522)( 96.8777, -115.9631)( 96.7642, -113.1441)
\qbezier(157.9528, -113.1441)(158.0666, -110.3151)(158.6301, -107.5763)
\qbezier( 96.7642, -113.1441)( 96.6503, -110.3151)( 96.0869, -107.5763)
\qbezier(158.6301, -107.5763)(159.2043, -104.7856)(160.2106, -102.2138)
\qbezier( 96.0869, -107.5763)( 95.5127, -104.7856)( 94.5064, -102.2138)
\qbezier(160.2106, -102.2138)(161.2561, -99.5422)(162.6943, -97.2428)
\qbezier( 94.5064, -102.2138)( 93.4609, -99.5422)( 92.0227, -97.2428)
\qbezier(162.6943, -97.2428)(164.2359, -94.7781)(166.0811, -92.9002)
\qbezier( 92.0227, -97.2428)( 90.4811, -94.7781)( 88.6359, -92.9002)
\qbezier(166.0811, -92.9002)(167.0860, -91.8775)(170.2582, -89.3842)
\qbezier( 88.6359, -92.9002)( 87.6310, -91.8775)( 84.4588, -89.3842)
\qbezier(170.2582, -89.3842)(172.4032, -87.6983)(174.5482, -86.0124)
\qbezier( 84.4588, -89.3842)( 82.3138, -87.6983)( 80.1688, -86.0124)
\qbezier(174.5482, -86.0124)(176.6931, -84.3265)(178.8381, -82.6406)
\qbezier( 80.1688, -86.0124)( 78.0238, -84.3265)( 75.8788, -82.6406)
\qbezier(178.8381, -82.6406)(180.9831, -80.9547)(183.1281, -79.2688)
\qbezier( 75.8788, -82.6406)( 73.7339, -80.9547)( 71.5889, -79.2688)
\qbezier(183.1281, -79.2688)(185.2731, -77.5829)(187.4181, -75.8970)
\qbezier( 71.5889, -79.2688)( 69.4439, -77.5829)( 67.2989, -75.8970)
\qbezier(187.4181, -75.8970)(189.5631, -74.2111)(191.7080, -72.5252)
\qbezier( 67.2989, -75.8970)( 65.1539, -74.2111)( 63.0089, -72.5252)
\qbezier(191.7080, -72.5252)(193.8530, -70.8392)(195.9980, -69.1533)
\qbezier( 63.0089, -72.5252)( 60.8640, -70.8392)( 58.7190, -69.1533)
\qbezier(195.9980, -69.1533)(198.1430, -67.4674)(200.2880, -65.7815)
\qbezier( 58.7190, -69.1533)( 56.5740, -67.4674)( 54.4290, -65.7815)
\qbezier(200.2880, -65.7815)(202.4330, -64.0956)(204.5780, -62.4097)
\qbezier( 54.4290, -65.7815)( 52.2840, -64.0956)( 50.1390, -62.4097)
\qbezier(204.5780, -62.4097)(206.7229, -60.7238)(208.8679, -59.0379)
\qbezier( 50.1390, -62.4097)( 47.9940, -60.7238)( 45.8491, -59.0379)
\qbezier(168.1132, -138.0271)(168.1132, -138.3810)(167.5574, -138.7301)
\qbezier(167.5574, -137.3242)(167.0015, -136.9751)(165.9050, -136.6404)
\qbezier(167.5574, -138.7301)(167.0015, -139.0792)(165.9050, -139.4139)
\qbezier(165.9050, -139.4139)(164.8085, -139.7486)(163.2012, -140.0599)
\qbezier(163.2012, -135.9944)(161.5939, -135.6831)(159.5197, -135.4039)
\qbezier(163.2012, -140.0599)(161.5939, -140.3711)(159.5197, -140.6504)
\qbezier(159.5197, -140.6504)(157.4455, -140.9297)(154.9609, -141.1694)
\qbezier(154.9609, -134.8849)(152.4763, -134.6452)(149.6492, -134.4516)
\qbezier(154.9609, -141.1694)(152.4763, -141.4091)(149.6492, -141.6026)
\qbezier(149.6492, -141.6026)(146.8221, -141.7962)(143.7295, -141.9384)
\qbezier(143.7295, -134.1159)(140.6369, -133.9738)(137.3632, -133.8869)
\qbezier(143.7295, -141.9384)(140.6369, -142.0805)(137.3632, -142.1674)
\qbezier(137.3632, -142.1674)(134.0895, -142.2543)(130.7240, -142.2835)
\qbezier(130.7240, -133.7708)(127.3585, -133.7415)(123.9930, -133.7708)
\qbezier(130.7240, -142.2835)(127.3585, -142.3127)(123.9930, -142.2835)
\qbezier(123.9930, -142.2835)(120.6275, -142.2543)(117.3538, -142.1674)
\qbezier(117.3538, -133.8869)(114.0801, -133.9738)(110.9875, -134.1159)
\qbezier(117.3538, -142.1674)(114.0801, -142.0805)(110.9875, -141.9384)
\qbezier(110.9875, -141.9384)(107.8949, -141.7962)(105.0678, -141.6026)
\qbezier(105.0678, -134.4516)(102.2406, -134.6452)( 99.7561, -134.8849)
\qbezier(105.0678, -141.6026)(102.2406, -141.4091)( 99.7561, -141.1694)
\qbezier( 99.7561, -141.1694)( 97.2715, -140.9297)( 95.1973, -140.6504)
\qbezier( 95.1973, -135.4039)( 93.1231, -135.6831)( 91.5158, -135.9944)
\qbezier( 95.1973, -140.6504)( 93.1231, -140.3711)( 91.5158, -140.0599)
\qbezier( 91.5158, -140.0599)( 89.9085, -139.7486)( 88.8120, -139.4139)
\qbezier( 88.8120, -136.6404)( 87.7155, -136.9751)( 87.1596, -137.3242)
\qbezier( 88.8120, -139.4139)( 87.7155, -139.0792)( 87.1596, -138.7301)
\qbezier( 87.1596, -138.7301)( 86.6038, -138.3810)( 86.6038, -138.0271)
\qbezier(157.9245, -114.5486)(157.9245, -114.8141)(157.5076, -115.0759)
\qbezier(157.5076, -114.0214)(157.0908, -113.7596)(156.2684, -113.5085)
\qbezier(157.5076, -115.0759)(157.0908, -115.3377)(156.2684, -115.5887)
\qbezier(156.2684, -115.5887)(155.4460, -115.8398)(154.2405, -116.0732)
\qbezier(154.2405, -113.0241)(153.0351, -112.7906)(151.4794, -112.5812)
\qbezier(154.2405, -116.0732)(153.0351, -116.3066)(151.4794, -116.5161)
\qbezier(151.4794, -116.5161)(149.9237, -116.7255)(148.0603, -116.9053)
\qbezier(148.0603, -112.1919)(146.1969, -112.0122)(144.0765, -111.8670)
\qbezier(148.0603, -116.9053)(146.1969, -117.0851)(144.0765, -117.2303)
\qbezier(144.0765, -117.2303)(141.9562, -117.3754)(139.6367, -117.4820)
\qbezier(139.6367, -111.6152)(137.3173, -111.5086)(134.8620, -111.4434)
\qbezier(139.6367, -117.4820)(137.3173, -117.5887)(134.8620, -117.6538)
\qbezier(134.8620, -117.6538)(132.4067, -117.7190)(129.8826, -117.7409)
\qbezier(129.8826, -111.3563)(127.3585, -111.3344)(124.8344, -111.3563)
\qbezier(129.8826, -117.7409)(127.3585, -117.7628)(124.8344, -117.7409)
\qbezier(124.8344, -117.7409)(122.3102, -117.7190)(119.8550, -117.6538)
\qbezier(119.8550, -111.4434)(117.3997, -111.5086)(115.0803, -111.6152)
\qbezier(119.8550, -117.6538)(117.3997, -117.5887)(115.0803, -117.4820)
\qbezier(115.0803, -117.4820)(112.7608, -117.3754)(110.6405, -117.2303)
\qbezier(110.6405, -111.8670)(108.5201, -112.0122)(106.6567, -112.1919)
\qbezier(110.6405, -117.2303)(108.5201, -117.0851)(106.6567, -116.9053)
\qbezier(106.6567, -116.9053)(104.7933, -116.7255)(103.2376, -116.5161)
\qbezier(103.2376, -112.5812)(101.6819, -112.7906)(100.4765, -113.0241)
\qbezier(103.2376, -116.5161)(101.6819, -116.3066)(100.4765, -116.0732)
\qbezier(100.4765, -116.0732)( 99.2710, -115.8398)( 98.4486, -115.5887)
\qbezier( 98.4486, -113.5085)( 97.6262, -113.7596)( 97.2093, -114.0214)
\qbezier( 98.4486, -115.5887)( 97.6262, -115.3377)( 97.2093, -115.0759)
\qbezier( 97.2093, -115.0759)( 96.7925, -114.8141)( 96.7925, -114.5486)
\qbezier(168.1132, -91.0701)(168.1132, -90.7162)(167.5574, -90.3671)
\qbezier(168.1132, -91.0701)(168.1132, -91.4240)(167.5574, -91.7731)
\qbezier(167.5574, -90.3671)(167.0015, -90.0181)(165.9050, -89.6833)
\qbezier(167.5574, -91.7731)(167.0015, -92.1222)(165.9050, -92.4569)
\qbezier(165.9050, -89.6833)(164.8085, -89.3486)(163.2012, -89.0374)
\qbezier(165.9050, -92.4569)(164.8085, -92.7916)(163.2012, -93.1029)
\qbezier(163.2012, -89.0374)(161.5939, -88.7261)(159.5197, -88.4468)
\qbezier(163.2012, -93.1029)(161.5939, -93.4141)(159.5197, -93.6934)
\qbezier(159.5197, -88.4468)(157.4455, -88.1676)(154.9609, -87.9279)
\qbezier(159.5197, -93.6934)(157.4455, -93.9727)(154.9609, -94.2124)
\qbezier(154.9609, -87.9279)(152.4763, -87.6882)(149.6492, -87.4946)
\qbezier(154.9609, -94.2124)(152.4763, -94.4520)(149.6492, -94.6456)
\qbezier(149.6492, -87.4946)(146.8221, -87.3010)(143.7295, -87.1589)
\qbezier(149.6492, -94.6456)(146.8221, -94.8392)(143.7295, -94.9813)
\qbezier(143.7295, -87.1589)(140.6369, -87.0167)(137.3632, -86.9298)
\qbezier(143.7295, -94.9813)(140.6369, -95.1235)(137.3632, -95.2104)
\qbezier(137.3632, -86.9298)(134.0895, -86.8430)(130.7240, -86.8137)
\qbezier(137.3632, -95.2104)(134.0895, -95.2973)(130.7240, -95.3265)
\qbezier(130.7240, -86.8137)(127.3585, -86.7845)(123.9930, -86.8137)
\qbezier(130.7240, -95.3265)(127.3585, -95.3557)(123.9930, -95.3265)
\qbezier(123.9930, -86.8137)(120.6275, -86.8430)(117.3538, -86.9298)
\qbezier(123.9930, -95.3265)(120.6275, -95.2973)(117.3538, -95.2104)
\qbezier(117.3538, -86.9298)(114.0801, -87.0167)(110.9875, -87.1589)
\qbezier(117.3538, -95.2104)(114.0801, -95.1235)(110.9875, -94.9813)
\qbezier(110.9875, -87.1589)(107.8949, -87.3010)(105.0678, -87.4946)
\qbezier(110.9875, -94.9813)(107.8949, -94.8392)(105.0678, -94.6456)
\qbezier(105.0678, -87.4946)(102.2406, -87.6882)( 99.7561, -87.9279)
\qbezier(105.0678, -94.6456)(102.2406, -94.4520)( 99.7561, -94.2124)
\qbezier( 99.7561, -87.9279)( 97.2715, -88.1676)( 95.1973, -88.4468)
\qbezier( 99.7561, -94.2124)( 97.2715, -93.9727)( 95.1973, -93.6934)
\qbezier( 95.1973, -88.4468)( 93.1231, -88.7261)( 91.5158, -89.0374)
\qbezier( 95.1973, -93.6934)( 93.1231, -93.4141)( 91.5158, -93.1029)
\qbezier( 91.5158, -89.0374)( 89.9085, -89.3486)( 88.8120, -89.6833)
\qbezier( 91.5158, -93.1029)( 89.9085, -92.7916)( 88.8120, -92.4569)
\qbezier( 88.8120, -89.6833)( 87.7155, -90.0181)( 87.1596, -90.3671)
\qbezier( 88.8120, -92.4569)( 87.7155, -92.1222)( 87.1596, -91.7731)
\qbezier( 87.1596, -90.3671)( 86.6038, -90.7162)( 86.6038, -91.0701)
\qbezier( 87.1596, -91.7731)( 86.6038, -91.4240)( 86.6038, -91.0701)
\qbezier(208.8679, -59.0379)(208.8679, -58.3301)(207.7562, -57.6320)
\qbezier(208.8679, -59.0379)(208.8679, -59.7457)(207.7562, -60.4439)
\qbezier(207.7562, -57.6320)(206.6446, -56.9338)(204.4515, -56.2644)
\qbezier(207.7562, -60.4439)(206.6446, -61.1420)(204.4515, -61.8115)
\qbezier(204.4515, -56.2644)(202.2585, -55.5949)(199.0439, -54.9724)
\qbezier(204.4515, -61.8115)(202.2585, -62.4809)(199.0439, -63.1034)
\qbezier(199.0439, -54.9724)(195.8293, -54.3499)(191.6809, -53.7914)
\qbezier(199.0439, -63.1034)(195.8293, -63.7259)(191.6809, -64.2845)
\qbezier(191.6809, -53.7914)(187.5325, -53.2328)(182.5633, -52.7534)
\qbezier(191.6809, -64.2845)(187.5325, -64.8430)(182.5633, -65.3224)
\qbezier(182.5633, -52.7534)(177.5942, -52.2740)(171.9399, -51.8869)
\qbezier(182.5633, -65.3224)(177.5942, -65.8018)(171.9399, -66.1889)
\qbezier(171.9399, -51.8869)(166.2856, -51.4998)(160.1005, -51.2155)
\qbezier(171.9399, -66.1889)(166.2856, -66.5761)(160.1005, -66.8604)
\qbezier(160.1005, -51.2155)(153.9153, -50.9311)(147.3679, -50.7574)
\qbezier(160.1005, -66.8604)(153.9153, -67.1447)(147.3679, -67.3185)
\qbezier(147.3679, -50.7574)(140.8205, -50.5836)(134.0895, -50.5252)
\qbezier(147.3679, -67.3185)(140.8205, -67.4922)(134.0895, -67.5507)
\qbezier(134.0895, -50.5252)(127.3585, -50.4667)(120.6275, -50.5252)
\qbezier(134.0895, -67.5507)(127.3585, -67.6091)(120.6275, -67.5507)
\qbezier(120.6275, -50.5252)(113.8965, -50.5836)(107.3491, -50.7574)
\qbezier(120.6275, -67.5507)(113.8965, -67.4922)(107.3491, -67.3185)
\qbezier(107.3491, -50.7574)(100.8017, -50.9311)( 94.6165, -51.2155)
\qbezier(107.3491, -67.3185)(100.8017, -67.1447)( 94.6165, -66.8604)
\qbezier( 94.6165, -51.2155)( 88.4313, -51.4998)( 82.7771, -51.8869)
\qbezier( 94.6165, -66.8604)( 88.4313, -66.5761)( 82.7771, -66.1889)
\qbezier( 82.7771, -51.8869)( 77.1228, -52.2740)( 72.1537, -52.7534)
\qbezier( 82.7771, -66.1889)( 77.1228, -65.8018)( 72.1537, -65.3224)
\qbezier( 72.1537, -52.7534)( 67.1845, -53.2328)( 63.0361, -53.7914)
\qbezier( 72.1537, -65.3224)( 67.1845, -64.8430)( 63.0361, -64.2845)
\qbezier( 63.0361, -53.7914)( 58.8877, -54.3499)( 55.6731, -54.9724)
\qbezier( 63.0361, -64.2845)( 58.8877, -63.7259)( 55.6731, -63.1034)
\qbezier( 55.6731, -54.9724)( 52.4585, -55.5949)( 50.2655, -56.2644)
\qbezier( 55.6731, -63.1034)( 52.4585, -62.4809)( 50.2655, -61.8115)
\qbezier( 50.2655, -56.2644)( 48.0724, -56.9338)( 46.9607, -57.6320)
\qbezier( 50.2655, -61.8115)( 48.0724, -61.1420)( 46.9607, -60.4439)
\qbezier( 46.9607, -57.6320)( 45.8491, -58.3301)( 45.8491, -59.0379)
\qbezier( 46.9607, -60.4439)( 45.8491, -59.7457)( 45.8491, -59.0379)
\put( 50.9434, -114.5486){\line(1, 0){ 47.6639}}
\put(156.1097, -114.5486){\vector(1, 0){ 47.6639}}
\put(102.4408, -114.5486){\line(1, 0){  3.8335}}
\put(110.1078, -114.5486){\line(1, 0){  3.8335}}
\put(117.7748, -114.5486){\line(1, 0){  3.8335}}
\put(125.4417, -114.5486){\line(1, 0){  3.8335}}
\put(133.1087, -114.5486){\line(1, 0){  3.8335}}
\put(140.7757, -114.5486){\line(1, 0){  3.8335}}
\put(148.4427, -114.5486){\line(1, 0){  3.8335}}
\put(127.3585, -167.9356){\line(0, 1){ 25.3588}}
\put(127.3585, -67.3011){\vector(0, 1){ 46.1797}}
\put(127.3585, -140.5748){\line(0, 1){  4.0040}}
\put(127.3585, -132.5667){\line(0, 1){  4.0040}}
\put(127.3585, -124.5587){\line(0, 1){  4.0040}}
\put(127.3585, -116.5506){\line(0, 1){  4.0040}}
\put(127.3585, -108.5426){\line(0, 1){  4.0040}}
\put(127.3585, -100.5345){\line(0, 1){  4.0040}}
\put(127.3585, -92.5265){\line(0, 1){  4.0040}}
\put(127.3585, -84.5184){\line(0, 1){  4.0040}}
\put(127.3585, -76.5104){\line(0, 1){  4.0040}}
\put(118.0741, -117.7919){\vector(-3, -1){ 41.6590}}
\qbezier(125.6392, -115.1492)(127.3585, -114.5486)(129.0778, -113.9480)
\qbezier(132.5165, -112.7468)(134.2358, -112.1462)(135.9552, -111.5456)
\qbezier(139.3939, -110.3444)(141.1132, -109.7438)(142.8325, -109.1432)
\qbezier(146.2712, -107.9420)(147.9906, -107.3414)(149.7099, -106.7408)
\qbezier(153.1486, -105.5396)(154.8679, -104.9390)(156.5873, -104.3384)
\qbezier(121.8566, -116.4706)(122.8882, -116.1102)(123.9198, -115.7498)
\put(160.0259, -103.1372){\line(3, 1){ 20.8231}}
\put( 68.7736, -134.5687){\makebox(0,0)[]{\normalsize{$x$}}}
\put(213.9623, -118.5526){\makebox(0,0)[]{\normalsize{$y$}}}
\put(119.7170, -19.7867){\makebox(0,0)[]{\normalsize{$\hat{z}$}}}
\end{picture}
\vspace{3mm} \newline
\hspace*{3mm} \parbox[t]{150mm}{Fig.1.
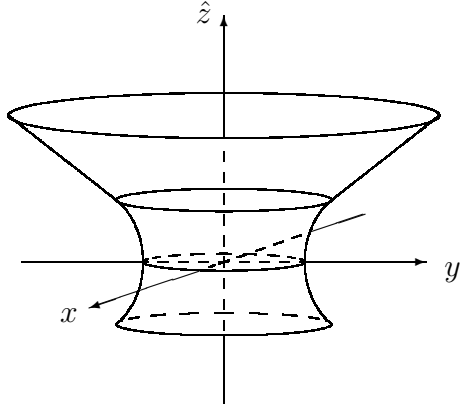
The $r \phi$-surface of the $3$-space $\hat{t} = constant$ as embedded
in Euclidean $3$-space. The region $\hat{z} < 0$ corresponds to
$0 < r < a \,$. Note that in the embedding diagram, $r = 0$ is
represented by the circle at the bottom of the surface.}
\vspace{12mm}
\makebox[8mm]{}
The expression (77) shows that the total line density $\lambda$
must satisfy $0 < \lambda < 2 {\lambda}_0$ in order that the embedding
shall be well defined.
\itm One may wonder why the embedded surface is bounded below at a
value of $\hat{z}$ corresponding to $r = 0$. The reason is that the value
$r = 0$ must be excluded from the region with the gas of strings because
of the spinning string at the axis.
\end{list}
\vspace{6mm}
{\bf 7. The Sagnac effect}
\vspace{-5mm} \newline
\begin{list}{}{\setlength{\leftmargin}{0mm}}
\item The local vorticity of a velocity field is given by eq.(3).
Applied to our solutions of Einstein's field equations in section 5
this formula gives a vanishing vorticity. One may then wonder whether
the existence of closed timelike geodesics is independent of the state
of rotation of the space. This would, however, be rather surprising,
since closed timelike curves have been shown to exist in some rotating
systems, i.e. in the G\"{o}del universe [1], outside a rotating
cosmic string [12], and in a rotating distribution of dust [8],[9].
\itm Usually the Sagnac effect is thought of as an effect appearing when
the apparatus is fixed in a rotating reference frame. In section 2 we have
argued that if $l > 0$, the frame $R$ is rotating, but that $R$ is not
rotating if $l < 0$. We shall now investigate whether an expriment with
the apparatus fixed in $R$ will show a Sagnac effect independent of the
sign of $l$. For light moving in a circular path with $ds = dr = dz = 0$,
the 4-velocity identity takes the form (including here the velocity of
light $c$)
\b
- c^2 f \dot{t}^2 + l \dot{\phi}^2 + 2 k \hs{0.5mm} \dot{t} \dot{\phi} = 0
\mbox{ .}
\e
Introducing the coordinate velocity $d \phi / dt$ we then get
\b
(\frl{d \phi}{dt})^2 + \frl{2kc}{l} \frl{d \phi}{dt}
- f \frl{c^2}{l} = 0
\e
The solutions are
\b
\left( \frl{d \phi}{dt} \right)_{\pm} = - \frl{k \pm D}{l} \hs{1.2mm} c
\e
We shall now consider light emitted in opposite directions along a
circle around the axis. Integrating eq.(80) around the circle we find the
travelling times for the two light signals
\b
t_1 = \frl{2 \pi l}{c (D + k)}
\mbox{\hspace{3mm} , \hspace{3mm}}
t_2 = \frl{2 \pi l}{c (D - k)}
\e
The time difference for light travelling around a circle in opposite
directions is
\b
\Delta t = t_2 - t_1 = \frl{4 \pi k}{cf}
\e
where we have used that $D^2 = k^2 + fl$.
The Sagnac effect is due to this time difference. In our models we
have $l < 0$. Then it follows from eq.(81) that $t_1 < 0$ and
$t_2 > 0$. Hence the light emitted in the negative $\phi$-direction
moves backwards in time. This implies the possibility that free
material particles can move along closed timelike worldlines.
\itm  As an illustation we may draw the light cones of light moving
in the circular paths. For comparison, we first consider the usual
case $l > 0$. Then $t_1$ and $t_2$ are both positive. The shape of
the light cone is as shown in Fig.2a. In our models with $l < 0$
the shape of the light cone is as shown in Fig.2b. In eq.(78)
we see that the first two terms are negative, which implies the
condition
\b
\dot{t} \dot{\phi} > 0
\e
This shows that $\dot{t}$ has opposite signs for light moving in
opposite directions. Hence there exist worldlines of light moving
in the negative time direction.
\vspace{55mm} \newline
\begin{picture}(50,0)(16,-162)
\qbezier(202.2926, -108.3870)(202.7814, -107.5477)(200.4947, -105.1805)
\qbezier(200.4947, -105.1805)(198.2080, -102.8133)(193.3936, -99.1748)
\qbezier(193.3936, -99.1748)(188.5792, -95.5362)(181.7588, -91.0206)
\qbezier(181.7588, -91.0206)(174.9385, -86.5050)(166.8512, -81.6016)
\qbezier(166.8512, -81.6016)(158.7639, -76.6982)(150.2861, -71.9384)
\qbezier(150.2861, -71.9384)(141.8083, -67.1787)(133.8587, -63.0783)
\qbezier(133.8587, -63.0783)(125.9091, -58.9780)(119.3492, -55.9814)
\qbezier(119.3492, -55.9814)(112.7892, -52.9848)(108.3298, -51.4167)
\qbezier(108.3298, -51.4167)(103.8704, -49.8485)(101.9947, -49.8788)
\qbezier(101.9947, -49.8788)(100.1190, -49.9091)(101.0304, -51.5345)
\qbezier(101.0304, -51.5345)(101.9418, -53.1599)(105.5414, -56.2043)
\qbezier(105.5414, -56.2043)(109.1411, -59.2487)(115.0389, -63.3821)
\qbezier(115.0389, -63.3821)(120.9368, -67.5156)(128.4937, -72.2902)
\qbezier(128.4937, -72.2902)(136.0506, -77.0648)(144.4476, -81.9632)
\qbezier(144.4476, -81.9632)(152.8447, -86.8616)(161.1719, -91.3529)
\qbezier(161.1719, -91.3529)(169.4992, -95.8442)(176.8543, -99.4418)
\qbezier(176.8543, -99.4418)(184.2094, -103.0393)(189.7953, -105.3532)
\qbezier(189.7953, -105.3532)(195.3811, -107.6671)(198.5925, -108.4467)
\qbezier(198.5925, -108.4467)(201.8038, -109.2263)(202.2926, -108.3870)
\qbezier(127.3585, -120.8973)(164.8256, -114.6422)(202.2926, -108.3870)
\qbezier(127.3585, -120.8973)(114.0895, -85.6040)(100.8206, -50.3106)
\put( 50.9434, -120.8973){\vector(1, 0){152.8302}}
\put(127.3585, -171.3986){\line(0, 1){ 50.5012}}
\put(127.3585, -71.6587){\vector(0, 1){ 39.1384}}
\put(127.3585, -117.1098){\line(0, 1){  3.7876}}
\put(127.3585, -109.5346){\line(0, 1){  3.7876}}
\put(127.3585, -101.9594){\line(0, 1){  3.7876}}
\put(127.3585, -94.3842){\line(0, 1){  3.7876}}
\put(127.3585, -86.8090){\line(0, 1){  3.7876}}
\put(127.3585, -79.2339){\line(0, 1){  3.7876}}
\put(127.3585, -120.8973){\vector(-3, -1){ 50.9434}}
\qbezier(131.1792, -119.6348)(133.0896, -119.0036)(135.0000, -118.3723)
\qbezier(138.8208, -117.1098)(140.7311, -116.4785)(142.6415, -115.8472)
\qbezier(146.4623, -114.5847)(148.3726, -113.9534)(150.2830, -113.3222)
\qbezier(154.1038, -112.0596)(156.0142, -111.4284)(157.9245, -110.7971)
\qbezier(161.7453, -109.5346)(163.6557, -108.9033)(165.5660, -108.2720)
\qbezier(169.3868, -107.0095)(171.2972, -106.3783)(173.2075, -105.7470)
\put( 68.7736, -139.8353){\makebox(0,0)[]{\normalsize{$x$}}}
\put(213.9623, -124.6849){\makebox(0,0)[]{\normalsize{$y$}}}
\put(119.7170, -31.2577){\makebox(0,0)[]{\normalsize{$t$}}}
\end{picture}
\begin{picture}(50,0)(-146,-162)
\qbezier(197.4244, -136.3283)(198.1074, -135.5213)(196.1104, -132.5595)
\qbezier(196.1104, -132.5595)(194.1133, -129.5977)(189.6525, -124.8021)
\qbezier(189.6525, -124.8021)(185.1918, -120.0064)(178.7507, -113.8965)
\qbezier(178.7507, -113.8965)(172.3097, -107.7866)(164.5863, -101.0246)
\qbezier(164.5863, -101.0246)(156.8629, -94.2626)(148.6942, -87.5813)
\qbezier(148.6942, -87.5813)(140.5254, -80.8999)(132.7966, -75.0233)
\qbezier(132.7966, -75.0233)(125.0677, -69.1467)(118.6162, -64.7115)
\qbezier(118.6162, -64.7115)(112.1647, -60.2764)(107.6897, -57.7634)
\qbezier(107.6897, -57.7634)(103.2147, -55.2504)(101.2012, -54.9319)
\qbezier(101.2012, -54.9319)( 99.1876, -54.6133)( 99.8537, -56.5238)
\qbezier( 99.8537, -56.5238)(100.5198, -58.4342)(103.7934, -62.3666)
\qbezier(103.7934, -62.3666)(107.0669, -66.2989)(112.5932, -71.8271)
\qbezier(112.5932, -71.8271)(118.1195, -77.3553)(125.2996, -83.8802)
\qbezier(125.2996, -83.8802)(132.4798, -90.4052)(140.5357, -97.2198)
\qbezier(140.5357, -97.2198)(148.5916, -104.0344)(156.6503, -110.4002)
\qbezier(156.6503, -110.4002)(164.7091, -116.7660)(171.8973, -121.9932)
\qbezier(171.8973, -121.9932)(179.0855, -127.2203)(184.6243, -130.7424)
\qbezier(184.6243, -130.7424)(190.1631, -134.2645)(193.4522, -135.6999)
\qbezier(193.4522, -135.6999)(196.7413, -137.1352)(197.4244, -136.3283)
\qbezier(127.3585, -120.8973)(162.3914, -128.6128)(197.4244, -136.3283)
\qbezier(127.3585, -120.8973)(113.6103, -88.0359)( 99.8621, -55.1744)
\put( 50.9434, -120.8973){\line(1, 0){ 76.4151}}
\put(170.4469, -120.8973){\vector(1, 0){ 33.3267}}
\put(131.1920, -120.8973){\line(1, 0){  3.8335}}
\put(138.8590, -120.8973){\line(1, 0){  3.8335}}
\put(146.5259, -120.8973){\line(1, 0){  3.8335}}
\put(154.1929, -120.8973){\line(1, 0){  3.8335}}
\put(161.8599, -120.8973){\line(1, 0){  3.8335}}
\put(127.3585, -171.3986){\line(0, 1){ 50.5012}}
\put(127.3585, -85.7834){\vector(0, 1){ 53.2631}}
\put(127.3585, -117.0971){\line(0, 1){  3.8002}}
\put(127.3585, -109.4967){\line(0, 1){  3.8002}}
\put(127.3585, -101.8963){\line(0, 1){  3.8002}}
\put(127.3585, -94.2958){\line(0, 1){  3.8002}}
\put(127.3585, -120.8973){\vector(-3, -1){ 50.9434}}
\qbezier(131.1792, -119.6348)(133.0896, -119.0036)(135.0000, -118.3723)
\qbezier(138.8208, -117.1098)(140.7311, -116.4785)(142.6415, -115.8472)
\qbezier(146.4623, -114.5847)(148.3726, -113.9534)(150.2830, -113.3222)
\qbezier(154.1038, -112.0596)(156.0142, -111.4284)(157.9245, -110.7971)
\qbezier(161.7453, -109.5346)(163.6557, -108.9033)(165.5660, -108.2720)
\put(170.8387, -106.5298){\line(3, 1){ 12.0481}}
\put( 68.7736, -139.8353){\makebox(0,0)[]{\normalsize{$x$}}}
\put(213.9623, -124.6849){\makebox(0,0)[]{\normalsize{$y$}}}
\put(119.7170, -31.2577){\makebox(0,0)[]{\normalsize{$t$}}}
\end{picture}
\vspace{10mm} \newline
\hspace*{9mm} \parbox[t]{75mm}{Fig.2a.
The light cone when $l > 0 \,$.}
\parbox[t]{80mm}{Fig.2b.
The light cone when $l < 0 \,$.}
\vspace{0mm} \newline
\itm Delgado, Schleich and S\"{u}ssmann [22] have introduced an angular
velocity ${\omega}_S$ characterising a global rotation of space, utilizing
the Sagnac effect. They compared the Sagnac effect in the G\"{o}del
universe with that in a rotating reference frame in flat spacetime in
the weak field approximation. In this approximation
\b
\Delta t_{RF} = - 4 \pi \frl{r^2 {\omega}_S}{c^2}
\e
in the rotating frame. This gives
\b
{\omega}_S = \frl{k}{r^2 f}
\e
By using eq.(42) we get for our solutions
\b
{\omega}_S = \frl{4GJ}{c^2 r^2}
\e
According to this interpretation space has a global rotation. However,
the angular velocity decreases with the distance from the axis in such
a way that the local vorticity vanishes. Hence the existence of a
non-vanishing global rotation seems to be decisive for the existence
of closed timelike geodesics in this spacetime.
\itm We have seen in section 2, however, that $R$ can only be interpreted
as a rotating frame when $l > 0$. When $l < 0$ the Sagnac effect is not
due to a rotation of the reference frame, but is a consequence of the fact
that the clocks in $R$ (showing $t$) are not Einstein synchronized. As we
noted, they can not be so, in order that the time $t$ shall not have a gap
around a closed circle about the axis. Consequently, the velocity of light
is anisotropic as measured with these clocks. This implies a non-vanishing
Sagnac effect although the apparatus which is fixed in $R$ does not rotate.
\end{list}
\vspace{6mm}
{\bf 8. Inertial dragging}
\vspace{-5mm} \newline
\begin{list}{}{\setlength{\leftmargin}{0mm}}
\item
%
%
In section 2 we showed that if $l > 0$, inertial frames will be dragged
around the axis in the reference frame $R$ with an angular velocity
$\omega = - k / l$. In this case there is therefore an inertial dragging
effect.
\itm We now want to investigate the case $l < 0$.
%
%
The conserved energy of the particle is
\b
p_{t} = \frl{\partial L}{\partial \dot{t}} =
-f \dot{t} + k \dot{\phi}
\e
Solving eqs.(6) and (87) with respect to $\dot{t}$ and $\dot{\phi} \,$,
we obtain
\b
\dot{t} = \frl{k p_{\phi} - l p_t}{D^2}
\mbox{\hspace{3mm} , \hspace{3mm}}
\dot{\phi} = \frl{f p_{\phi} + k p_t}{D^2}
\e
The $4$-velocity identity of the particle is
\b
- f \dot{t}^2 + l \dot{\phi}^2 + 2 k \hs{0.5mm} \dot{t} \dot{\phi}
+ e^{\mu} \dot{r}^2 = \epsilon
\mbox{ ,}
\e
where $\epsilon = -1,0,1$ respectively for a timelike, lightlike or
spacelike interval in spacetime.
Inserting the expressions (88) for $\dot{t}$ and $\dot{\phi} \,$ leads to
\b
f p_{\phi}^2 + 2k p_{\phi} p_{t} - l p_{t}^2
= (\epsilon - e^{\mu} \dot{r}^2) D^2
\e
If $p_{\phi} = 0$, then
\b
l {p_{\hs{0.4mm} t}}^{\hs{-1.4mm} 2} = (- \epsilon + e^{\mu} \dot{r}^2) D^2
\e
Hence $l < 0$ implies that $\epsilon = 1$. Consequently local inertial
frames having $p_{\phi} = 0$, must follow spacelike worldlines in
spacetime when $l < 0$. They then have a tachyon-like character.
This is not to be interpreted as some sort of super dragging effect.
One must rather conclude that when $l < 0$ no inertial frames with
velocity $v < c$ exist. This is not due to rotation, but to the way
that the clocks in $R$ must be synchronised in order that the time $t$
showed by them shall not have a gap when one passes around a circle
about the axis. This property of the spacetime is presumably due to the
spinning cosmic string at the axis.
\end{list}
\vspace{6mm}
{\bf 9. Conclusion}
\vspace{-5mm} \newline
\begin{list}{}{\setlength{\leftmargin}{0mm}}
\item In the present paper we have found a new class of solutions of
Einstein's field equations. They represent cylindrically symmetric
spacetimes in a region of limited radial extension with a non rotating
gas of vacuum energy and strings, and with a spinning cosmic string along
the symmetry axis. Outside this region there is empty spacetime.
\itm Timelike circular curves are characterized by the four velocity
identity $\vec{u} \cdot \vec{u} = -1 \,$. Considering a circular timelike
curve in the plane $z = constant \,$, this equation takes the form (14).
Surprisingly, such a timelike curve may fullfill the condition
$\dot{t} = 0 \,$, which means that the curve is closed in spacetime.
This requires $l < 0$ in the line element (1). If such a curve shall
also be a geodesic, the requirements (19) must also be met.
\itm We have investigated two solutions of the class mentioned above,
one with constant pressure, and one in which the pressure is given by
eq.(55), both containing closed timelike geodesics. In these models it
is possible for a free particle to travel along such
a curve and come back to the event it started from.
\itm Rosa and Letelier [23] have shown that the closed timelike geodesics
followed by such free particles are not stable. However, in the present
case the force needed to move along a closed timelike curve perturbing the
geodesic may be made arbitrarily small.
\end{list}
\vspace{6mm}
{\bf References}
\begin{list}{}{\setlength{\leftmargin}{0mm}}
\item 1. Kurt G\"{o}del, {\it An Example of a New Type of Cosmological
Solutions of Einstein's Field Equations of Gravitation}, Rev.Mod.Phys.
{\bf 21}, 447 (1949).
\item 2. W.J. van Stockum, Pros.R.Soc.Edinb. {\bf 57}, 135 (1937).
\item 3. F.J.Tipler, {\it Rotating cylinders and the possibility of global
causality violation}, Phys.Rev. {\bf D9}, 2203 (1974).
\item 4. J.R.Gott, {\it Closed Timelike Curves Produced by Pairs of
Moving Cosmic Strings: Exact Solutions}, Phys.Rev.Lett. {\bf 66},
1126 (1991).
\item 5. S.DeDeo and J.R.Gott, {\it Eternal time machine
$(2 + 1)$-dimensional anti-de Sitter space}, Phys.Rev. {\bf D66}, 084020
(2002).
\item 6. P.Collas and D.Klein, {\it Causality Violating Geodesics in
Bonnor'r Rotating Dust Metric}, Gen.Rel.Grav. {\bf 36}, 2549 (2004).
\item 7. I.D.Soares, {\it Inhomogeneous rotating universes with closed
timelike geodesics of matter}, J.Math.Phys. {\bf 21}, 521 (1980).
\item 8. B.R.Steadman, {\it Causality Violation on van Stockum Geodesics},
Gen.Rel.Grav. {\bf 35}, 1721 (2003).
\item 9. W.B.Bonnor and B.R.Steadman, {\it Exact solutions of the
Einstein Maxwell equations with closed timelike curves.},
Gen.Rel.Grav. {\bf 37}, 1833 (2005).
\item 10. V.M.Rosa and P.L.Letelier, {\it Stability of Closed Timelike
Geodesics.}, arXiv:gr-qc/0703148.
\item 11. V.M.Rosa and P.L.Letelier, {\it Spinning Strings, Black Holes and
Stable Closed Timelike Geodesics.}, arXiv:gr-qc/0704.1109v2.
\item 12. S.Deser and R.Jackiw, {\it Time Travel ?}, Comments
Nucl.Part.Phys. {\bf 20}, 337 (1992) (arXiv:hep-th/9206094).
\item 13. J.N.Islam, {\it Rotating fields in general relativity},
Cambridge University Press (1985), page 21.
\item 14. W.Rindler and V.Perlick, {\it Rotating Coordinates as Tools for
Calculating Circular Geodesics and Gyroscopic Precession}, General
Relativity and Gravitation, {\bf 22}, 1067 (1990).
\item 15. L.Herrera, F.M.Paiva and N.O.Santos, {\it Gyroscope Precession
in Cylindrically Symmetric Spacetimes}, Class.Quant.Grav. {\bf 17}, 1549
(2000) (arXiv:gr-qc/0001075).
\item 16. M.J.Weyssenhoff {\it Metrisches Feld und Gravitationsfeld},
Bull.Acad.Polonaise, Sc.Lett. Ser.A, 252 (1937).
\item 17. L.D.Landau and E.M.Lifshitz, The Classical Theory of Fields,
Pergamon Press, 1971, paragraph 88, page 248.
\item 18. J.D.Barrow and C.G.Tsagas, {\it Dynamics and stability of the
G\"{o}del universe}, gr-qc/0308067.
\item 19. F.I.Cooperstock and S.Tieu, {\it Closed Timelike Curves and
Time Travel: Dispelling the Myth}, Found.Phys. {\bf 35}, 1497 (2005).
\item 20. H.Culetu, {\it On the Deser-Jackiw Spinning String Spacetime},
arXiv:hep-th/0602014.
\item 21. A.Vilenkin and E.P.S. Shellard, Cosmic Strings and Other
Topological Defects, Cambridge University Press (1994), page 192.
\item 22. A.Delgado, W.P.Schleich and G.S\"{u}ssmann, {\it Quantum
gyroscopes and G\"{o}del's universe: entanglement opens a new testing
ground for cosmology}, New Journal of Physics {\bf 4}, 37.1 (2002).
\item 23. V.M.Rosa and P.S.Letelier, {\it Stability of Closed Timelike
Geodesics in different Spacetimes}, arXiv:gr-qc/0706.3212.
\end{list}
\vspace{6mm}
{\bf Acknowledgements}
\begin{list}{}{\setlength{\leftmargin}{0mm}}
\item We want to thank the referees for useful and inspiring comments
leading to a substantial improvement of the manuscript.
\end{list}
\end{document}